\RequirePackage{fix-cm}
\documentclass[final,a4paper,twocolumn,dvipdfmx]{svjour3}

\usepackage{amsmath,amssymb}
\usepackage{graphicx}
\usepackage[sort&compress,numbers]{natbib}
\usepackage{bm}
\newcommand{\pd}[2]{\frac{\partial#1}{\partial#2}}

\newcommand{\sgn}{\text{sgn}}
\newcommand{\diag}{\text{diag}}
\newcommand{\rank}{\text{rank}\,}
\renewcommand{\Re}{\text{Re}\,}
\renewcommand{\Im}{\text{Im}\,}
\newcommand{\leqn}[1]{\label{eqn:#1}}
\newcommand{\reqn}[1]{(\ref{eqn:#1})}
\newcommand{\lfig}[1]{\label{fig:#1}}
\newcommand{\rfig}[1]{Fig.\;\ref{fig:#1}}
\newcommand{\Rfig}[1]{Figure\;\ref{fig:#1}}
\newcommand{\ltab}[1]{\label{tab:#1}}
\newcommand{\rtab}[1]{Table\,\ref{tab:#1}}
\newcommand{\rtabs}[2]{Tables\,\ref{tab:#1}\,and\,\ref{tab:#2}}
\newcommand{\lsec}[1]{\label{sec:#1}}
\newcommand{\rsec}[1]{Section\ \ref{sec:#1}}
\newcommand{\dbm}[1]{\dot{\bm{#1}}}

\newcommand{\mm}{{\mathcal M}}
\newcommand{\clD}{{\mathcal D}}
\newcommand{\step}{\chi}
\newcommand{\Dth}{\eta}
\newcommand{\Lam}{\Lambda}
\newcommand{\lam}{\lambda}
\newcommand{\lammax}{\lambda_{\max}}
\newcommand{\argmax}{\mathop{\text{arg\,max}}}
\newcommand{\xx}{{\bm x}}
\newcommand{\dxx}{{\dbm x}}
\newcommand{\ff}{{\bm f}}
\newcommand{\torq}{{\bm \tau}}

\spaceskip=\fontdimen2\font plus 1.4\fontdimen3\font minus 1.5\fontdimen4\font
\usepackage{url}

\makeatletter
\def\makeheadbox{{%
\hbox to0pt{\vbox{\baselineskip=10dd\hrule\hbox
to\hsize{\vrule\kern3pt\vbox{\kern3pt
\hbox{This is an article published in {\bfseries\@journalname}.}
\hbox{The final publication is available at Springer via 
 \;\url{http://dx.doi.org/10.1007/s11071-015-2376-7}}
\kern3pt}\hfil\kern3pt\vrule}\hrule}%
\hss}}}
\makeatother

\begin{document}
\journalname{Nonlinear Dynamics}
\title{Nonlinear analysis on purely mechanical 
 stabilization of a wheeled inverted pendulum on a slope}
 \author{Katsutoshi Yoshida \and
         Munehisa Sekikawa \and
         Kenta Hosomi}
 \institute{K. Yoshida \and M.Sekikawa\at
               Department of Mechanical and Intelligent Engineering,
	       Utsunomiya University,
	       7--1--2 Yoto, Utsunomiya, Tochigi 321--8585, Japan
	       \email{yoshidak@katzlab.jp}
	   \and
	   K. Hosomi \at
	      Isuzu Motors Limited, 6--26--1 Minami, Shinagawa,
 	      Tokyo 140--8722, Japan
 }
 \date{Received: 22 May 2015 / Accepted: 31 August 2015}
 \maketitle

 \begin{abstract}
  This paper investigates the potential for stabilizing an inverted
  pendulum without electric devices, using gravitational potential
  energy. We propose a wheel\-ed mechanism on a slope, specifically, a
  wheeled double pendulum, whose second pendulum transforms gravity
  force into braking force that acts on the wheel.  In this paper, we
  derive steady-state equations of this system and conduct nonlinear
  analysis to obtain parameter conditions under which the standing
  position of the first pendulum becomes asymptotically stable.
  In this asymptotically stable condition, the proposed mechanism
  descends the slope in a stable standing position, while dissipating
  gravitational potential energy via the brake mechanism.
  By numerically continuing the stability limits in the parameter space,
  we find that the stable parameter region is simply connected. This
  implies that the proposed mechanism can be robust against errors in
  parameter setting.
 \end{abstract}

\section{Introduction}

Electric and electronic control devices are indispensable for a variety
of modern technologies. However, these technologies typically become
useless during massive power outages such as those caused by natural or
other disasters.
In this paper, we consider a non-electrified alternative control
method to stabilize an inverted pendulum using gravitational
potential. 
Our proposed mechanism is a wheeled inverted pendulum that descends a
slope. The brake mechanism of our proposed mechanism transforms gravity
force into friction force between the pendulum and the wheel. This
friction produces a restoring force by which the pendulum is
asymptotically stabilized in a standing position.

Approaches similar to ours can be found in the field of passive dynamic
walking, pioneered by McGeer \cite{McGeer1990}, in which two-legged
mechanisms are designed to stably walk down a slope that consume only
gravitational potential energy. Extensive studies have been reported on
the passive dynamic walking, including experimental development of
passive walkers \cite{McGeer1990,Coleman1998,Ikemata2009} and nonlinear
analyses of passive dynamic walking based on simplified models
\cite{Goswami1998,Garcia1998,HIRATA2011}.  
Early insights into the use of such passivity can also be found in the
study of passive gravity-gradient attitude stabilization
\cite{BLACK1964,FISCHELL1964,He1999,Peiffer2000} wherein the alignment
of one axis of a satellite along the earth's local vertical direction was
achieved without the use of active control elements.

On the other hand, the wheeled inverted pendulum has attracted
significant attention in the fields of control engineering and
robotics. Because of the applications of wheeled inverted pendulums in
personal mobility devices, including the
Segway$^{\text{\textregistered}}$ \cite{Ulrich2005}, methods for
controlling wheeled inverted pendulums have been developed via
approaches such as partial feedback linearization \cite{Pathak2005},
inclined surface control \cite{Kim2006}, sliding-mode velocity control
\cite{Huang2010}, neuro-fuzzy-based control \cite{Su2010}, and robust
control based on a quasi-linear parameter-varying model
\cite{Vermeiren2011}.  Not surprisingly, these studies implied the use
of electric devices. 

In this paper, we merge concepts from passive dynamic walking and
studies of the wheeled inverted pendulum to derive our new mechanical
design for the non-electrified stabilization of a wheeled inverted
pendulum. 
As stated above, we propose a wheeled double pendulum mechanism, whose
second pendulum transforms gravity force into braking force that acts on
the wheel.
To investigate the dynamic stabilities of this newly proposed
mechanism,
we start with deriving a nonlinear analytical model of the
mechanism to examine the stabilities of its steady states. Three types
of critical points arise in the analytical model.  These critical points
are analytically characterized and numerically continued in the
parameter space to obtain stability limits for the steady standing
motions.  It is found that the stability of the proposed mechanism is
limited by Hopf bifurcation and vanishing external resistance on the
wheel.

\section{Wheeled inverted pendulum with friction control}

\begin{figure}[t]
 \centering
 \includegraphics[width=\hsize]{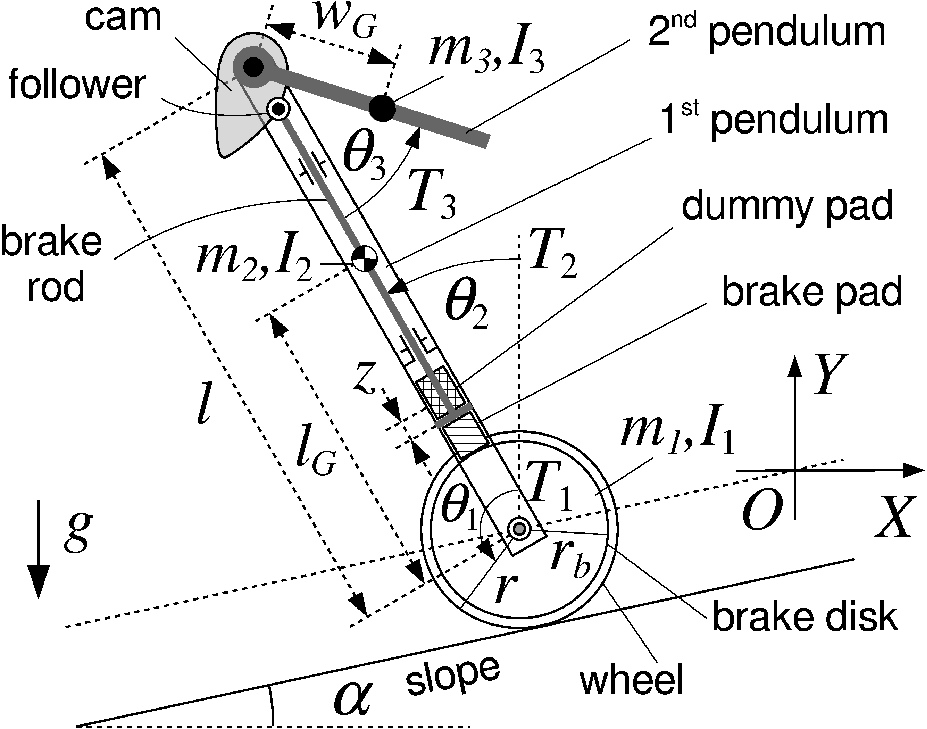} 
 \caption{Friction-controlled wheeled inverted pendulum (FCWIP).}
 \lfig{model}
\end{figure}

We propose a wheeled inverted pendulum with friction control, as shown
in \rfig{model}, that comprises 1) a wheel placed on a slope without
slipping, 2) a double pendulum suspended on the wheel axis, and 3) a
friction control mechanism that generates a braking force on the wheel
proportional to the angle between the first and second pendulums.
Hereinafter, we refer to this model as a friction-controlled wheeled
inverted pendulum (FCWIP).

Configuration of the FCWIP can be described by a three-dimensional
generalized coordinate: ($^T$ denotes transpose) 
\begin{equation}
 \bm{\theta}=(\theta_1,\theta_2,\theta_3)^T,
\end{equation}
where $\theta_1$ is the rotational angle of the wheel, $\theta_2$ is the
absolute slant angle of the first pendulum, and $\theta_3$ is the relative
angle of the second pendulum from the first pendulum.
In addition, we consider the corresponding generalized force:
\begin{equation}
 \bm{T}=(T_1,T_2,T_3)^T,
  \leqn{T}
\end{equation}
where $T_i$ is a torque acting on $\theta_i$.

\subsection{Wheeled double pendulum}

Unless the friction control mechanism (or $\bm T$) is specified, the
FCWIP in \rfig{model} is simply a wheeled double pendulum whose
Lagrangian is given by
\begin{align}
 L &:= T - U: \\
 T &:= 
 \frac{1}{2} {\dot\theta}_1^2 (Q_1 r+I_1)
 +\frac{1}{2} {\dot\theta}_2^2 \{-2 Q_3 l \cos (\theta _3)+Q_2+I_2\}
\notag\\%
 &\qquad+\frac{1}{2} {\dot\theta}_3^2 (Q_3 w_G+I_3)
 +{\dot\theta}_2 {\dot\theta}_3 Q_3 \{w_G-l \cos (\theta _3)\}
\notag\\%
 &\quad+{\dot\theta}_1 {\dot\theta}_2 \{
   -Q_3 r \cos (\alpha-\theta _2-\theta _3)
   +Q_4 r \cos (\alpha -\theta _2)
 \}
\notag\\
&\qquad
 -{\dot\theta}_1 {\dot\theta}_3 Q_3 r \cos (\alpha-\theta _2-\theta _3),
\\
 U &:=  -g \{Q_3 \cos (\theta _2+\theta _3)-Q_4 \cos (\theta _2)
 +\theta _1 Q_1 \sin (\alpha )\}
 \leqn{U}
 ,
\end{align}
with
\begin{align}
  Q_1 &= (m_1+m_2+m_3)r,
  \notag\\
  Q_2 &= m_2 l_G^2 + m_3 (w_G^2+l^2),
  \notag\\
  Q_3 &= m_3 w_G,
  \quad
  Q_4 = m_2 l_G+ m_3l,
\end{align}
where the physical parameters are listed in \rtab{para1}.

\begin{table}[t]
 \centering
 \caption{Mechanical parameters of the wheeled double pendulum.}
 \newcommand{\mltwo}[1]{\multicolumn{2}{@{}l}{#1}}
 \begin{tabular}{c@{\quad}p{46mm}l}\hline
  \multicolumn{2}{l}{Parameters} & \multicolumn{1}{l}{Values}
  \\\hline
  $m_1$ & mass of wheel & 0.1 kg
	  \\
  $m_2$ & mass of 1st pendulum & 0.2 kg
	  \\
  $m_3$ & mass of 2nd pendulum & 1 kg
	  \\
  $I_1$ & moment of inertia of wheel  & $m_1r^2/2$ kg$\cdot$m$^2$
	  \\
  $I_2$ & moment of inertia of 1st pendulum  & $m_2l^2/12$ kg$\cdot$m$^2$
	  \\
  $I_3$ & moment of inertia of 2nd pendulum  & $0.25$ kg$\cdot$m$^2$
	  \\
  $r$ & radius of wheel & 0.2 m
	  \\
  $l$ & length of 1st pendulum & 1 m
	  \\
  $l_G$ & placement of center of gravity of 1st pendulum & $l/2$
	  \\
  $w$ & length of 2nd pendulum & 0.7 m
	  \\
  $w_G$ & placement of center of gravity of 2nd pendulum & 0.6 m
	  \\
  $g$ & acceleration of gravity & 9.8 m/s$^2$
	  \\
  $\alpha$ & angle of slope & 0.1 rad
	  \\
  \hline
 \end{tabular}
 \ltab{para1}
\end{table}

On substituting $L$ into Lagrange's equations with the generalized
force $\bm T$, we obtain the equations of the motion of the wheeled
double pendulum as
\begin{equation}
 M\ddot{\bm{\theta}} = \bm{F} + \bm{T},
 \leqn{eom}
\end{equation}
with
\begin{align}
 &M^T=M,\; M_{11} = Q_1 r+I_1
  ,\notag\\
 &M_{12} = -Q_3 r \cos (\alpha -\theta _2-\theta _3) + Q_4 r \cos (\alpha -\theta _2)
  ,\notag\\
 & M_{13} = -Q_3r \cos (\alpha -\theta _2-\theta _3) 
  ,\notag\\
 & M_{22} = -2 Q_3 l \cos (\theta _3) +Q_2 + I_2
  ,\notag\\
 & M_{23} = Q_3 \{w_G-l \cos (\theta _3)\}
  ,\quad
 M_{33} = Q_3 w_G + I_3
 ,
 \leqn{MM}
\end{align}
and
\begin{align}
  F_1=
  & Q_3 r ({\dot\theta}_2+{\dot\theta}_3){}^2 \sin (\alpha -\theta _2-\theta _3)
  \notag\\&\qquad\qquad
   -Q_4 r {\dot\theta}_2^2 \sin (\alpha -\theta _2)+g Q_1 \sin (\alpha ), \notag\\
  F_2=
  &- Q_3l {\dot\theta}_3 (2 {\dot\theta}_2+{\dot\theta}_3) \sin (\theta _3)
  \notag\\&\qquad\qquad
  -g\{ Q_3 \sin (\theta _2+\theta _3)- Q_4 \sin (\theta _2)\}, \notag\\
  F_3=
  & Q_3l {\dot\theta}_2^2 \sin (\theta _3)-g Q_3 \sin (\theta _2+\theta _3)
 ,
 \leqn{Fi}
\end{align}
where $M_{ij}$ represents the $(i,j)$ component of the matrix $M$ and
$F_i$ is the $i$th component of the vector $\bm F$.

\subsection{Friction control mechanism}

Next, we introduce a friction control mechanism (FCM) into the wheeled
double pendulum by specifying $\bm T$ as follows.

Let $z$ be a displacement of the brake rod outputted from the cam
mechanism, as shown in \rfig{model}, and suppose that the cam function
$z(\theta_3)$ is given as a linear function:
\begin{equation}
 z=z(\theta_3):= \rho\,(\theta_3-\Dth)
  ,\quad \dot z = \rho\,\dot\theta_3
  \leqn{cam}
\end{equation}
where $\rho$ is a cam ratio and $\Dth>0$ is an offset angle.
Accordingly, the follower is expected to follow both the positive and
negative rotation of the cam.

\begin{figure}[t]
 \centering
 \includegraphics[width=\hsize]{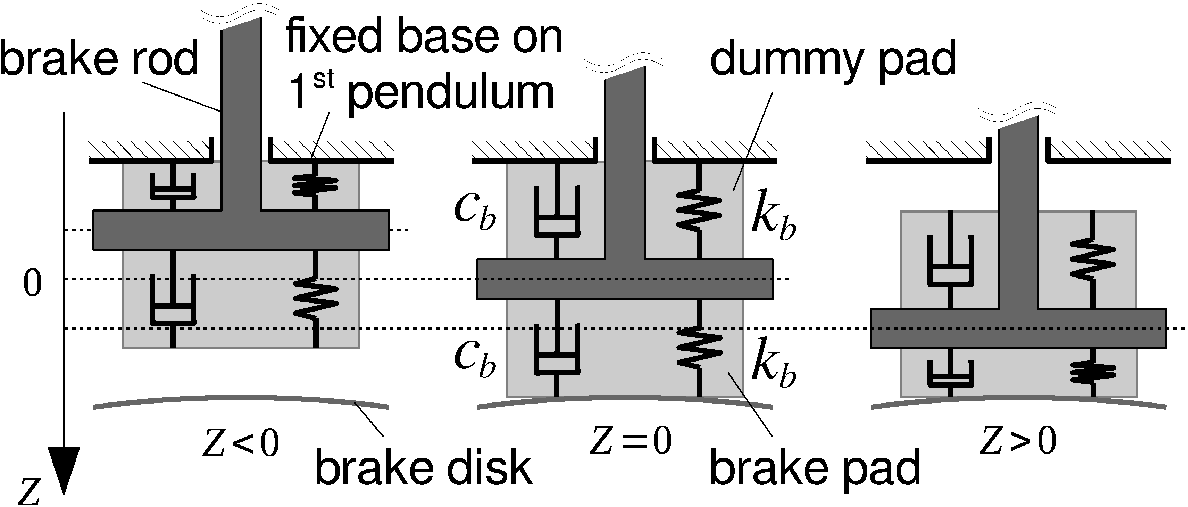} 
 \caption{Example of the structure of the brake mechanism}
 \lfig{FCM}
\end{figure}

Then, we consider a brake mechanism, as shown in \rfig{FCM}. In this
mechanism, a pad (in light gray) is bonded on a brake rod (in dark gray)
and sandwiched between the fixed base on the first pendulum and the
brake disk without clearance at $z=0$.
We refer to the lower half of the pad as a brake pad and the upper half as a
dummy.
The dummy pad has no function in terms of braking but is assumed to have the same
mechanical property as the brake pad.

Thus, the brake pad touches the brake disk when $z\geq0$ but is
separated from the disk when $z<0$.  We assume that the brake rod receives a
continuous reaction force from the pads in the following form:
\begin{align}
 R = R(z,\dot z) :=& -(k_b z + c_b \dot z)
  \notag\\
  =& -\rho\left\{k_b (\theta_3-\Dth) + c_b \dot \theta_3\right\}
  ,
\end{align}
where $c_b$ and $k_b$ are viscoelastic coefficients of the pad.  

The reaction force $R$ produces a torque on $\theta_3$ as a generalized
force $F_{\theta_3}$ on $\theta_3$, given by
\begin{equation}
 F_{\theta_3} = \pd{z}{\theta_3}R(z,\dot z) = \rho R(z,\dot z).
\end{equation}
At the same time, we assume that $R$ causes a Coulomb friction force
between the brake pad and the brake disk. This force can be modeled by a
tangential force on the contact surface as
\begin{equation}
 F_R = \mu R(z,\dot z)\,\sgn(\dot\theta_1-\dot\theta_2)\step(z)
  \leqn{FR0}
\end{equation}
where $\mu$ is the Coulomb friction coefficient, $\sgn(\cdot)$ is the unit
signum function, and $\step(\cdot)$ is the unit step function representing
the separation of the brake pad from the brake disk.
We have the torques $T_i$ on $\theta_i$ ($i=1,2,3$) as
\begin{equation}
 \begin{cases}
  T_1 = r_b F_R  - c_1|\dot\theta_1|\dot\theta_1,
  \\
  T_2 = -r_bF_R,
  \\
  T_3 = F_{\theta_3} = \rho R(z,\dot z),
 \end{cases}
 \leqn{torque0}
\end{equation}
where $c_1$ is the coefficient of the quadratic resistance including
aerodynamic force on the wheel (or $\theta_1$).  \rtab{para2} summarizes
the parameters of the FCM and quadratic resistance.
Note that the value of the spring coefficient listed in
\rtab{para2} can be obtained approximately from a medium-carbon steel
rod (Young's modulus 205 GPa) of $5\times10^{-4}$ m diameter and 0.5 m
length.

\begin{table}[bt]
 \centering 
 \caption{Parameters of the FCM.}
 \begin{tabular}{c@{\quad}p{45mm}l}\hline
  \multicolumn{2}{l}{Parameters} & \multicolumn{1}{l}{Values}
  \\\hline
  $r_b$ & radius of brake disk & $0.18$ m
          \\
  $\rho$ & cam ratio & 1/20
	  \\
  $k_b$ & spring coefficient of the brake & $8\times 10^{4}$ N/m
	  \\
  \hline
  $c_b$ & viscous coefficient of the brake & ($2\times10^4$ Ns/m)
	  \\
  $\mu$ & Coulomb friction coefficient of the brake & ($0.249$)
	  \\
  $\Dth$ & offset of the brake mechanism & ($2\times10^{-4}$ rad)
	  \\
  $c_1$ & coefficient of quadratic resistance on the wheel
      & ($5\times10^{-4}$ Nms$^2$)
	  \\
  \hline
 \end{tabular}
 \\Parentheses around values denote nominal values.\hfill
 \ltab{para2}
\end{table}

Therefore, we derive the dynamic model of the FCWIP as the wheeled
double pendulum in \reqn{eom} with the braking torque in
\reqn{torque0}.

\subsection{Numerical examples}

\begin{figure}[t]
 \centering
 \includegraphics[width=\hsize, trim=0 0 0 0]{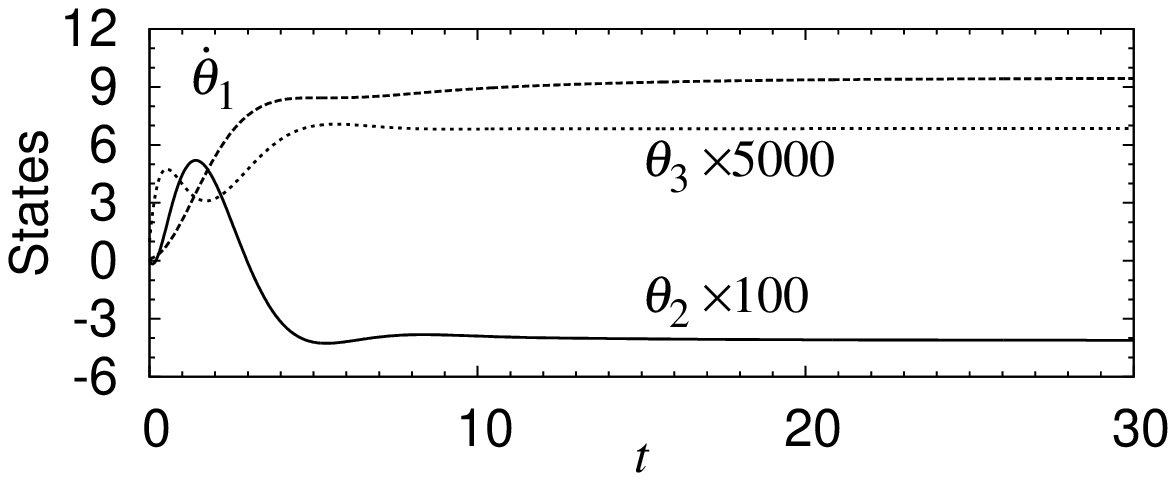}
 \caption{Time responses of the FCWIP for the condition listed in
 \rtabs{para1}{para2}.}
 \lfig{res}
 \vskip\floatsep
 \includegraphics[width=\hsize, trim=0 10 0 0]{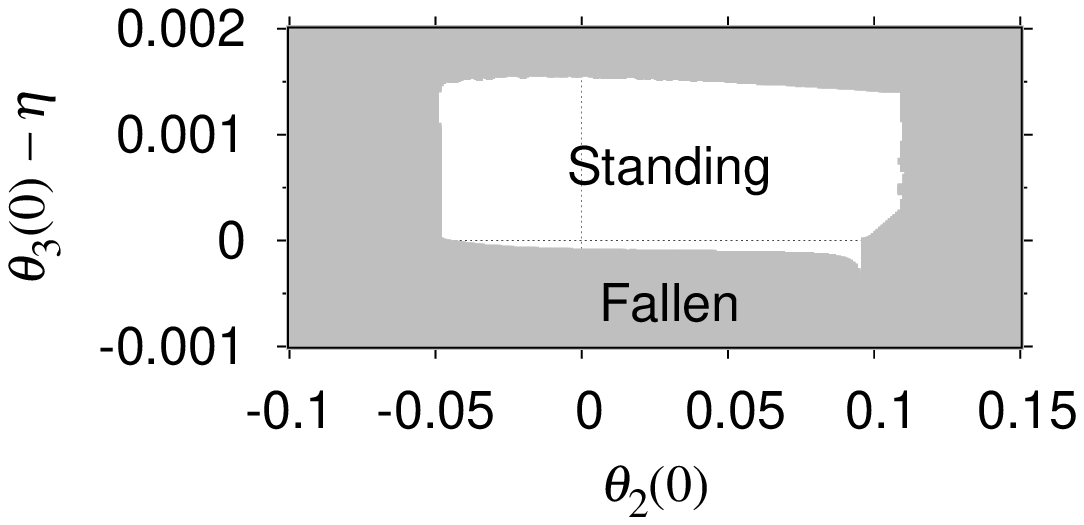}
 \caption{Attraction basin of the stable steady state in \rfig{res}.}
 \lfig{basin}
\end{figure}

\Rfig{res} shows a numerical solution of the FCWIP model obtained by
solving \reqn{T}, \reqn{eom}, and \reqn{torque0} from the trivial
initial state $\theta_1(0)=\theta_2(0)=0$, $\theta_3(0)=\Dth$ (or
$z(0)=0$), and $\dot\theta_i(0)=0$ ($i=1,2,3$).
The parameter values are listed in \rtab{para1} and \ref{tab:para2} that are empirically chosen to achieve a stable standing motion.
For numerical integration, a fourth-order Runge-Kutta-Gill method is
employed with a time step of $10^{-3}$.

As shown in \rfig{res}, the FCWIP model becomes asymptotically stable
under the suitable parameter condition. In this example, the angle of the
first pendulum $\theta_2(t)$ converges to a small negative value 
representing a standing position that is slightly slanted toward the upside of the
slope. Consequently, the angle of the second pendulum $\theta_3(t)$
converges to a small positive value that represents it hanging from
the first pendulum to produce the brake force $F_R$ in \reqn{FR0}.
Moreover, the descent velocity of the wheel $\dot\theta_1(t)$ converges
to $9.46$ rad/s ($6.81$ km/h in translational velocity).

\Rfig{basin} shows the sets of initial angles $\theta_2(0), \theta_3(0)$
of the first and second pendulums, respectively. The other initial
values are set as $\theta_1(0)=\dot\theta_i(0)=0$ $(i=1,2,3)$. The area
hatched in gray is the set of initial angles from which the state
converges to fallen positions of the FCWIP model, and the white area
surrounded by the hatched area is the set that converges to the steady
standing state, as shown in \rfig{res}.  From \rfig{basin}, it appears
that the set of the initial angles that belongs to the standing position
forms a mostly connected area.  Therefore, it can be expected that the
proposed FCWIP model exhibits some robustness against disturbance of the
initial angles.

Considering potential real-life applications of the proposed
system, dependencies of the parameters on the size of the attraction
basin that belongs to the standing position could be a crucial
problem. This will be addressed in a future study.

\section{Steady-state analysis}

\subsection{State-space representation}

For a simple expression, we perform a time scale transformation
$t:=qt^\ast$, where $q$ is a time scale and $t^\ast$ is nondimensional
time. Taking a state vector:
\begin{equation}
 \xx := 
  (\theta_1,\theta_2,\theta_3-\Dth,
  q\dot\theta_1,q\dot\theta_2,q\dot\theta_3)^T
  ,\; \dot\theta_i := d\theta_i/dt,
  \leqn{statevec}
\end{equation}
we transform the dynamic model \reqn{eom} and \reqn{torque0} to a
state-space form:
\begin{equation}
 \dxx = \mm(\xx)^{-1}\Big\{\ff(\xx) + \torq(\xx)\Big\},
  \; \dxx := d\xx/dt^\ast,
 \leqn{ssm}
\end{equation}
with
\begin{equation}
 \mm(\xx) := \diag\Big(E_3,M(\xx)\Big)
  ,\;
 \leqn{ssm:matrix}
\end{equation}
\begin{equation}
 \ff(\xx) :=
 \begin{pmatrix}
  x_4\\x_5\\x_6\\q^2\bm F(\xx)
 \end{pmatrix}
 ,\;
 \torq(\xx) :=
 \begin{pmatrix}
  0\\ 0\\ 0\\ q^2\bm T(\xx)
 \end{pmatrix}
 ,
 \leqn{ssm:vectors}
\end{equation}
where $M(\xx)$, $\bm F(\xx)$, and $\bm T(\xx)$ are the matrix and
vectors in the dynamic model \reqn{eom} and \reqn{torque0} via
\reqn{statevec}, and $E_3$ is a $3\times3$ identity matrix.

We choose $q:=(k_b\rho^2)^{-1/2}$ to normalize the spring coefficient
$k_b$ and introduce nondimensional parameters listed in
\rtab{para3}. In this case, the components of the vectors in
\reqn{ssm:vectors} are obtained as
\begin{equation}
\begin{cases}
  q^2F_1=
  -Q_3 r (x_5+x_6)^2 \sin (\Dth +x _2+x _3 -\alpha )
  \\\qquad\qquad
   +Q_4 r x_5^2 \sin (x_2 -\alpha)+g^\ast Q_1 \sin (\alpha ), \\
  q^2F_2=
  - Q_3l x_6 (2 x_5+x_6) \sin (\Dth+x_3)
  \\\qquad\qquad
  -g^\ast\{ Q_3 \sin (\Dth+x_2+x_3)- Q_4 \sin (x_2)\}, \\
   q^2F_3=
  Q_3l x_5^2 \sin (\Dth+x_3)-g^\ast Q_3 \sin (\Dth+x_2+x_3)
 ,
\end{cases}
 \leqn{qFi}
\end{equation}
and
\begin{equation}
\begin{cases}
  q^2T_1 =
 - \mu^\ast \sgn(x_4-x_5)\step(x_3)
 \left\{x_3+c_b^\ast x_6\right\}
  \\\qquad\qquad
 - c_1|x_4|x_4 
  ,\\
  q^2T_2 = \mu^\ast \sgn(x_4-x_5)\step(x_3)
 \left\{x_3+c_b^\ast x_6\right\}
  ,\\
  q^2T_3 = 
 -\left\{x_3+c_b^\ast x_6\right\}
 .
\end{cases}
 \leqn{qT}
\end{equation}

\begin{table}[t]
 \centering
 \caption{Nondimensional parameters.}
 \begin{tabular}{c@{\quad}p{34mm}ll}\hline
  \multicolumn{2}{l}{Parameters} 
  & \multicolumn{1}{l}{Values}
  & \multicolumn{1}{l}{Definition}
  \\\hline
  $q$ & time scale & $200^{-1/2}$ & $(k_b\rho^2)^{-1/2}$ 
          \\
  $k_b^{\ast}$ & spring coefficient & $1$ & $(q^2\rho^2)k_b$
	  \\
  $g^{\ast}$ & acceleration of gravity & $4.9\times 10^{-2}$ & $(q^2)g$
	  \\
  \hline
  $\mu^{\ast}$ & Coulomb friction coefficient & ($0.8964$) & $(r_b\rho^{-1})\mu$ 
	  \\
  $c_b^{\ast}$ & viscous coefficient & ($3.536$) & $(q\rho^2)c_b$
	  \\
  \hline
 \end{tabular}
 \begin{flushleft}\small
  The values are transformed from the original parameters in
  \rtabs{para1}{para2}. Parentheses around values denote nominal
  values.
 \end{flushleft}
 \ltab{para3}
\end{table}

\subsection{Assumption of steady state}
\lsec{assumption}

In view of the numerical example presented in \rfig{res}, we consider a steady
state $\xx^\ast$ of the FCWIP model \reqn{ssm} that satisfies
\begin{equation}
 \dxx  = \dxx^\ast:=\left(\omega,0,0,0,0,0\right)^T
  ,\quad
  \omega>0\;(\text{constant}).
  \leqn{SSdef}
\end{equation}
The steady state $\xx^\ast$ mentioned above describes %
the rotation of the wheel down the slope with a constant angular
velocity $x_4^\ast = \omega>0$ 
while the first and second pendulum maintain certain steady angles
$x_2^\ast$ and $x_3^\ast$, respectively, with $x_5^\ast=x_6^\ast=0$.
The angles $x_2^\ast$ and $x_3^\ast$ are assumed to satisfy the
following conditions:
\begin{enumerate}
 \item[(A)] $x_2^\ast < 0$: the first pendulum reaches a standing
       position (slightly) slanted to the upside of the slope.
 \item[(B)] $x_3^\ast > 0$ (or $z^\ast>0$): due to (A), the
       second pendulum hangs from the first pendulum (due to gravity and while
       maintaining $x_2^\ast <0$) to produce the brake force $F_R$ in
       \reqn{FR0}.
\end{enumerate}
These conditions are required to stabilize the first pendulum in a
standing position.
Because, they guarantee existence of the braking force $F_R$ caused by
a negative clearance between the brake pad and disk, $x_3^\ast>0$ (or
$z^\ast>0$), and that is mechanically caused by $x_2^\ast<0$.
Otherwise, the braking force vanishes, and the FCWIP becomes nothing more
than an uncontrolled wheeled double pendulum that can never be
stabilized around the standing position.

In addition, note that the FCWIP model, in absence of the floor model of
the slope, can theoretically have another stable steady state, a static
equilibrium where the second pendulum is hanging down at rest at
$x_2^\ast=\pi+\epsilon_2$ and $x_3^\ast=\epsilon_3$ for small
$\epsilon_2,\epsilon_3>0$.

\subsection{Steady-state equation}

The steady-state equation is given by
\begin{equation}
 \dxx^\ast = \mm(\xx^\ast)^{-1}\Big\{\ff(\xx^\ast) + \torq(\xx^\ast)\Big\},
 \leqn{SS0}
\end{equation}
where the derivative $\dxx^\ast$ is the constant vector already defined in
\reqn{SSdef} and $\xx^\ast$ is an unknown vector representing the steady
state.  Multiplying both the sides by $\mm(\xx^\ast)$, we obtain
\begin{equation}
 \ff(\xx^\ast) + \torq(\xx^\ast) = \mm(\xx^\ast)\dxx^\ast
  =\diag\big(E_3,M(\xx^\ast)\big) \dxx^\ast = \dxx^\ast
 \leqn{SS1}
\end{equation}
where the last equality is due to the zero components of
$\dxx^\ast:=(\omega,0,0,0,0,0)$.  Therefore, the steady-state
condition is obtained as follows:
\begin{align}
& x_4^\ast=\omega,\quad x_5^\ast=x_6^\ast = 0,
 \leqn{ss0}\\
& 0 = q^2F_1+q^2T_1 
 = g^\ast Q_1 \sin (\alpha )  - c_1\omega^2  - \mu^\ast  x_3^\ast,
 \leqn{ss1}\\
& 0 = q^2F_2+q^2T_2
 =-g^\ast\{ Q_3 \sin (\Dth+x_2^\ast+x_3^\ast)
  \notag\\&\qquad\qquad\qquad\qquad\qquad
- Q_4 \sin (x_2^\ast)\}
 +\mu^\ast  x_3^\ast,
 \leqn{ss2}\\
& 0 = q^2F_3+q^2T_3
 = -g^\ast Q_3 \sin (\Dth+x_2^\ast+x_3^\ast)
 - x_3^\ast,
 \leqn{ss3}
\end{align}
where $|x_4^\ast|x_4^\ast = \omega^2$ and
$\sgn(x_4^\ast-x_5^\ast)\step(x_3^\ast)=1$ are substituted according to
the assumption: $x_4^\ast-x_5^\ast =x_4^\ast=\omega>0$ and $x_3^\ast>0$
in \rsec{assumption}.

Note that the equations \reqn{ss1}, \reqn{ss2}, and \reqn{ss3}
represent the balance of the torque from the brake force
and the gravity force at about $\theta_1$, $\theta_2$, and
$\theta_3$, respectively. In particular, \reqn{ss1} can also be derived
from the balance of the energy supply from the gravitational
potential and the energy consumption via Coulomb friction and
quadratic resistance.

The steady-state equations in \reqn{ss0}, \reqn{ss1}, \reqn{ss2}, and
\reqn{ss3} can be reduced to the following form:
\begin{equation}
 \left\{
 \begin{aligned}
  &x_5^\ast=x_6^\ast = 0,
 \\
 &{c_1} (x_4^\ast)^2 
  = \relax{g^\ast Q_1 \sin (\alpha )- \mu^\ast  x_3^\ast}
 \quad > 0,
 \\
 &\sin (x_2^\ast) = -\frac{(1+\mu^\ast)x_3^\ast}{g^\ast Q_4},
 \\
 &\sin (x_2^\ast+x_3^\ast+\Dth) = -\frac{x_3^\ast}{g^\ast Q_3 }.
 \end{aligned}
 \right.
 \leqn{SSeqn}
\end{equation}
Thus, we have derived the steady-state equations of the FCWIP model
with three unknowns $x_2^\ast$, $x_3^\ast$, and $x_4^\ast$ ($=\omega$).
Note that the angle of the wheel in the steady state $
x_1^\ast(t)\propto \omega t$ never appears explicitly in these
steady-state equations.

It is clear from \reqn{SSeqn} that under the given mechanical structure
and environment, the nontrivial components of steady state $(x_2^\ast,$
$x_3^\ast,$ $x_4^\ast)$ depend on three parameters, namely $\mu^\ast$,
$\eta$, and $c_1$. More precisely, $x_2^\ast$ and $x_3^\ast$ can be
solved independently of $x_4^\ast$, and they depend on the
nondimensional friction $\mu^\ast$ and the offset $\eta$ of the
FCM. After that, $x_4^\ast$ is obtained as a function of $x_3^\ast$
depending on $c_1$.

\subsection{Jacobian matrix at steady state}
\lsec{linear}

We consider a variation $\delta\xx$ of $\xx$ around $\xx^\ast$ as $\xx
:= \xx^\ast + \delta\xx$ and substitute it into the state-space model
\reqn{ssm} as
\begin{math}
  \mm(\xx^\ast + \delta\xx)\{\dxx^\ast+\delta\dxx\} 
 = (\ff+\torq)(\xx^\ast + \delta\xx).
 \leqn{lssm0}
\end{math}
The $i$th component of the left side can be written by
the Einstein notation as
\begin{align}
 L_i&=\mm_{ij}(\xx^\ast+\delta\xx)
 \Big\{\dot x_j^\ast + \delta\dot x_j\Big\}
 \notag\\
 &=\Big\{\mm_{ij}(\xx^\ast)+\pd{\mm_{ij}}{x_k}\delta x_k\Big\}
 \Big\{\dot x_j^\ast + \delta\dot x_j\Big\}
 +O(\delta\xx,\delta\dot x_j)^2
 \notag\\
 &=
 \mm_{ij}(\xx^\ast)\dot x_j^\ast
 +\mm_{ij}(\xx^\ast)\delta\dot x_j
 \notag\\&\qquad\qquad
 +\pd{\mm_{ij}}{x_k}\delta x_k\dot x_j^\ast
 +O(\delta\xx,\delta\dxx)^2.
\end{align}
Due to the structures of $\mm=\diag(E_3,M)$ and
$\dxx^\ast=(\omega,0,0,0,0,0)^T$, we have
\begin{equation}
 \pd{\mm_{ij}}{x_k}\delta x_k\dot x_j^\ast
  = 0 =
  \begin{cases}
   0\cdot\delta x_k\omega & (j=1)\\
   \pd{\mm_{ij}}{x_k}\delta x_k\cdot0& (j>1)
  \end{cases}
\end{equation}
for all $i$ and $j$. 
Therefore, neglecting the second and higher order term of $\delta\xx$
and $\delta\dxx$, we arrive at a variation equation of \reqn{ssm} as
\begin{equation}
 \delta\dxx = \mm(\xx^\ast)^{-1}\Big\{
  D\ff(\xx^\ast) + D\torq(\xx^\ast)
  \Big\} \delta\xx
  =: J(\xx^\ast) \delta\xx
  ,
 \leqn{lssm}
\end{equation}
where $D\ff(\xx^\ast)$ denotes the Jacobian matrix of $\ff(\xx)$ around
$\xx^\ast$ and $J(\xx^\ast)$ provides a closed-loop state matrix whose
eigenvalues represent the stabilities of the FCWIP model.  The components of
$J(\xx^\ast)$ are given by
\newcommand{\gQCa}{g^\ast Q_3C_2^\ast}
\newcommand{\gQCb}{g^\ast Q_4C_1^\ast}
\begin{equation}
 D\ff(\xx^\ast) =
 \begin{pmatrix}
  0 & 0 & 0 & 1 & 0 & 0 \\
  0 & 0 & 0 & 0 & 1 & 0 \\
  0 & 0 & 0 & 0 & 0 & 1 \\
  0 & 0 & 0 & 0 & 0 & 0 \\
  0 & \gQCb-\gQCa & -\gQCa & 0 & 0 & 0 \\
  0 & -\gQCa            & -\gQCa & 0 & 0 & 0 \\
 \end{pmatrix},
 \leqn{Dff}
\end{equation}
where 
$C_1^\ast := \cos(x_2^\ast)$, 
$C_2^\ast := \cos(x_2^\ast+x_3^\ast+\Dth)$, and
\begin{equation}
 D\torq(\xx^\ast) =
 \begin{pmatrix}
  0 & 0 & 0 & 0 & 0 & 0 \\
  0 & 0 & 0 & 0 & 0 & 0 \\
  0 & 0 & 0 & 0 & 0 & 0 \\
  0 & 0 & -\mu^\ast & -2c_1x_4^\ast & 0 & -c_b^\ast\mu^\ast \\
  0 & 0 & \mu^\ast  & 0 & 0 & c_b^\ast\mu^\ast \\
  0 & 0 & -1             & 0 & 0 & -c_b^\ast \\
 \end{pmatrix}.
 \leqn{Dtt}
\end{equation}
To obtain \reqn{Dff} and \reqn{Dtt}, $|x_4|x_4 = x_4^2$ and
$\sgn(x_4-x_5)\step(x_3)=1$ are assumed, because $x_3(t) >0$, $x_4(t)
>0$, and $x_4(t)-x_5(t) >0$ hold for $x_i(t) = x_i^\ast + \delta x_i(t)$,
$|\delta x_i(t)|\ll 1$ $(i=1,\cdots,6)$.

The results in \reqn{Dff} and \reqn{Dtt} imply that the stability of the
FCWIP model depends on the nondimensional viscous coefficient $c_b^\ast$
in addition to the parameters $\mu^\ast$, $\eta$, and $c_1$ that determine
the steady states $(x_2^\ast,x_3^\ast,x_4^\ast)$.

In summary, we have found that under a given mechanical structure and environment,

\begin{itemize}
 \item the steady angles $(x_2^\ast,x_3^\ast)$ depend on $(\mu^\ast, \eta)$,
 \item the steady descent velocity $x_4^\ast$ depends on $(\mu^\ast, \eta, c_1)$, and
 \item the stability depends on $(\mu^\ast, \eta, c_1, c_b^\ast)$.
\end{itemize}

\subsection{Eigenvalue equation}

It can be proved that $\rank[J(\xx^\ast)] = 5<6=\dim\xx$, which follows from the assumption of the uniform motion $\dot x_1(t)=\omega$.  Thus, the characteristic equation of $J(\xx^\ast)$ is given
in the following form:
\begin{align}
 &\det\left(J(\xx^\ast)-\lam E_6\right) =  \lam\,h(\lam) = 0,
 \\
 &h(\lam)=\lam^5+a_1\lam^4+a_2\lam^3+a_3\lam^2+a_4\lam+a_5,
 \leqn{h}
\end{align}
where $\det(\cdot)$ denotes a determinant and $E_6$ is a $6\times6$ identity
matrix. For simplicity, we refer to $h(\lam)=0$ as an eigenvalue equation
of the FCWIP model.

Therefore, the steady state $\xx^\ast$ becomes stable if the maximal
real part of the eigenvalues is negative, that is
\begin{equation}
 \Lam:=\Re\lammax <0,\quad  \lammax:=\argmax_{\lam} \Re(\lam),
 \leqn{Lam}
\end{equation}
where $\lam_i$ ($i=1,\cdots,5$)
are the roots of $h(\lam)=0$.

\section{Critical points of steady state}

As mentioned in \rsec{linear}, the stabilities of the
steady state $\bm x^\ast$ depend on $\mu^\ast$, $\eta$, $c_1$, and
$c_b^\ast$.  Here, we provide some numerical examples of that dependency. %
Thus, the parameter values are set to those listed in
\rtab{para1}, \ref{tab:para2}, and \ref{tab:para3} by default unless
otherwise noted.

\subsection{Dependency on $\mu^\ast$ and $\Dth$}

We first examine the dependency on the nondimensional friction
$\mu^\ast$ and the offset $\Dth$,  which determine the steady angles
$x_2^\ast$ and $x_3^\ast$ via \reqn{SSeqn}.

\Rfig{eq-mu} shows the maximal real part of the eigenvalue $\Lam$ and
the nontrivial components $x_i^\ast$ $(i=2,3,4)$ of the steady state $\bm
x^\ast$ as functions of the nondimensional friction $\mu^\ast$. $\Lam$
and $x_i^\ast$ are obtained from numerical solutions of \reqn{SSeqn}
by Newton's method and \reqn{Lam}, respectively.
The solid line in the top graph represents $\Lam$, and the solid and
dotted lines in the bottom graph represent stable and unstable
$x_i^\ast$, respectively.  The values of $x_i^\ast$ $(i=2,3)$ are scaled
to share a common vertical axis.

\begin{figure}[t]
 \centering
 \includegraphics[width=\hsize]{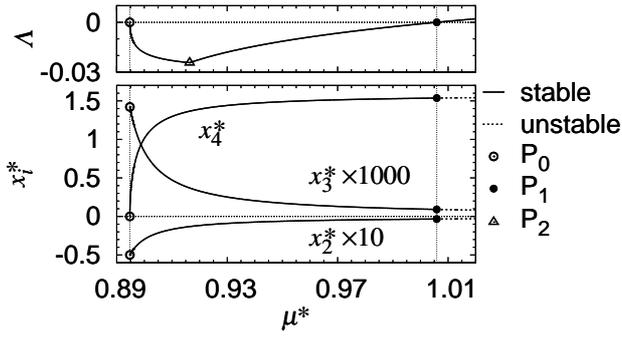}
 \caption{Maximal real part of eigenvalue $\Lam$ and steady state
 $x_i^\ast$ as functions of the nondimensional friction $\mu^\ast$.}
 \lfig{eq-mu}
\end{figure}

It is clear from \rfig{eq-mu} that three types of critical points
$P_0$, $P_1$, and $P_2$ appear, which are denoted as open circles,
filled circles, and a triangle, respectively.
$P_0$ gives an infimum $\inf_{\mu^\ast} (x_4^\ast)=0$ of
the descent velocity $x_4^\ast>0$, $P_1$ gives a stability boundary,
and $P_2$ gives a folded (nonsmooth) minimum of
$\Lam(\mu^\ast)$.

\begin{figure}[t]
 \centering
 \includegraphics[width=\hsize, trim=0 0 20 0]{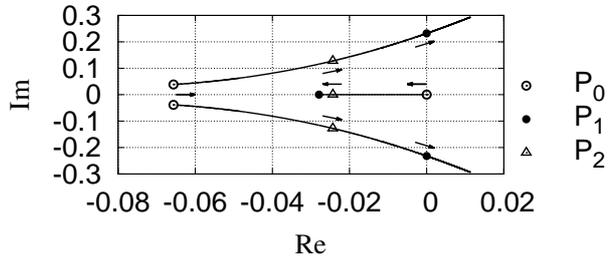}
 \caption{Root loci along the steady states in \rfig{eq-mu}.}
 \lfig{locus}
\end{figure}

To characterize these points, \rfig{locus} plots the root loci of
$h(\lam)=0$ in \reqn{h} along the steady states $\bm x^\ast$ in
\rfig{eq-mu}.
It can be numerically proven that under the considered condition, the
eigenvalue equation $h(\lam)=0$ has three real roots and a pair of
complex conjugate roots as
\begin{gather}
 \lam_0 = s_0,\; \lam_{1\pm}=s_1\pm j\Omega,\; \lam_2=s_2,\; \lam_3 = s_3,
  \notag\\ s_0, s_1 \gg s_2,s_3<0,
  \leqn{5eig}
\end{gather}
where $s_i$, $\Omega$ are real numbers and $j:=\sqrt{-1}$. Among the
five roots, only the first three roots $\lam_0$, $\lam_{1+}$, and
$\lam_{1-}$ affect the stability change, and only these three roots
appear in the range of \rfig{locus}.

On the basis of the root loci obtained in \rfig{locus}, we can
characterize the critical points in terms of eigenvalue types as
follows:
\begin{itemize}
 \item (P0): $P_0$ is a point such that $\lammax=0$.
 \item (P1): $P_1$ is a Hopf bifurcation point, for which 
             the loci cross the imaginary axis at
             $\lammax=\pm j\Omega$, where
	     $\Omega$ is an angular frequency of a limit cycle.
 \item (P2): $P_2$ is a point such that maximal real root $s_0$ and the
	     real part of the complex conjugate roots $s_1$ coincide, at
	     which point they switch roles to produce the maximal real
	     part.
\end{itemize}
Physically speaking, $P_0$ and $P_1$ provide stability limits, and
$P_2$ maximizes the total stability or minimizes the time constant of the
FCWIP model.
The descent velocity vanishes ($x_4^\ast=0$) at $P_0$ in this case, and
a self-excited oscillation (limit cycle) emerges for $\Lam(\mu^\ast)>0$
near $P_1$.

\begin{figure}[t]
 \centering
 \includegraphics[width=.75\hsize, trim=0 10 0 0]{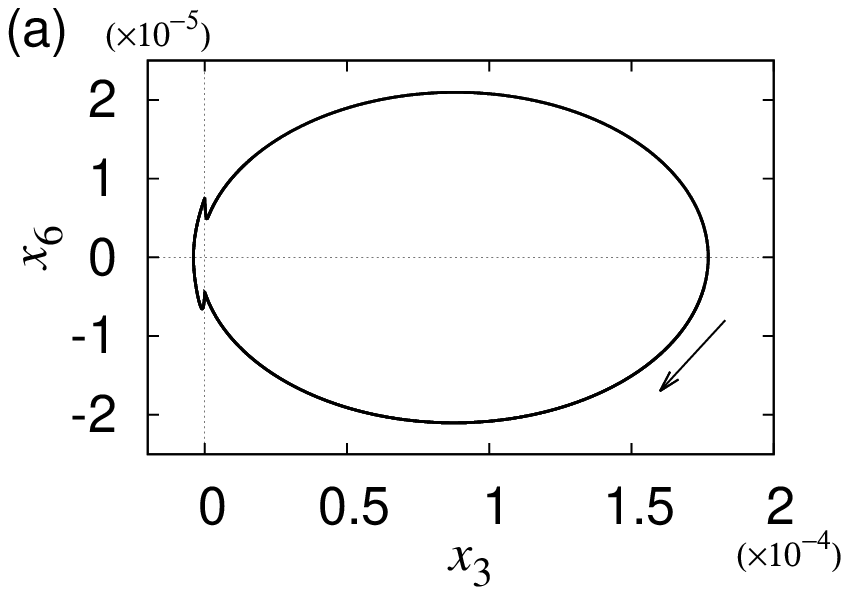}\par
 \includegraphics[width=.75\hsize, trim=0 10 0 0]{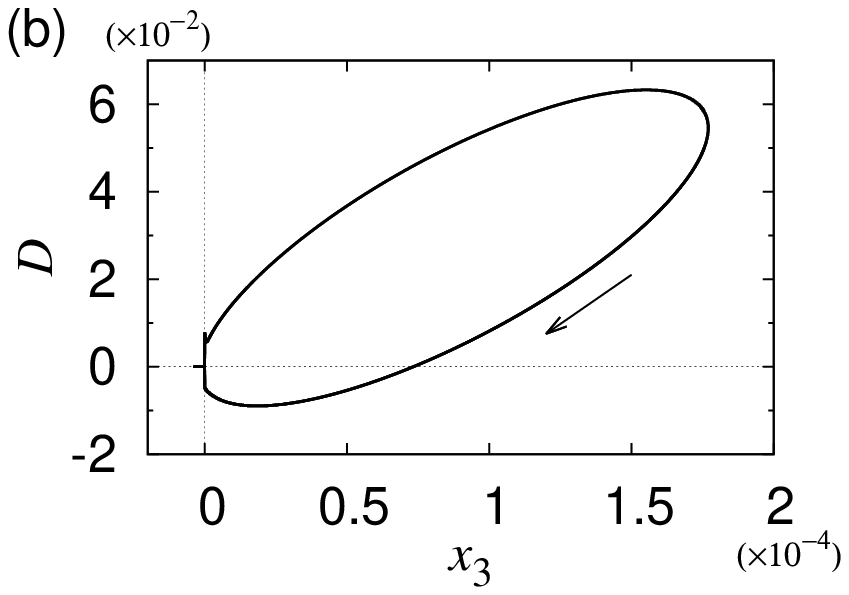}
 \caption{Limit cycle for $\mu^\ast=1.01$ (just after Hopf bifurcation).}
 \lfig{cycle}
\end{figure}

\Rfig{cycle} (a) shows the limit cycle for $\mu^\ast=1.01$ immediately after
the Hopf bifurcation point $P_1$. 
Under this condition, the FCWIP model descends the slope in a standing
position while the angles of the pendulums oscillate slightly and
periodically.
As this limit cycle is stable, $P_1$ is
identified as 
a Hopf bifurcation point.
In general, a limit cycle occurs because of the temporal presence of negative
resistance (i.e., negative energy consumption per unit time).
In our problem, this is given by the braking torque $T_2$ in \reqn{qT}
multiplied by the friction velocity $(x_4-x_5)$, specifically as
\begin{align}
 \clD &= T_2\cdot(x_4 - x_5)
 \notag\\
 &= q^{-2} \mu^\ast\sgn(x_4 - x_5)\step(x_3)(x_3+c_b^\ast x_6)
 \cdot(x_4 - x_5)
 \notag\\
 &= q^{-2} \mu^\ast|x_4 - x_5|\step(x_3)(x_3+c_b^\ast x_6)
 \notag\\
 &=: C\step(x_3)(x_3+c_b^\ast x_6)\quad(C>0),
\end{align}
where $\sgn(x)x=|x|$ is applied. As $\step(x_3)$ is a step function,
$\clD<0$ occurs when $x_3\geq0$ and $x_3+c_b^\ast x_6<0$. 
This implies that when $\clD<0$, the brake pad (see \rfig{FCM}
recalling $z=\rho x_3,\dot z=(\rho/q)x_6$) will be released with
a velocity less than $x_6<-x_3/c_b^\ast\leq0$ while the pad is still
pressed against the disk ($x_3\geq0$).
\Rfig{cycle} (b) shows the energy consumption $\clD$ along the limit
cycle as a function of $x_3$, which numerically confirms the presence of
$\clD<0$. 
During this negative energy consumption, the cycle comes
across the deadband border $x_3 = z = 0$. At this point, the step
function $\step(x_3)$ jumps from 1 to 0. This causes the $\clD$ to jump
from a negative to a positive value, similar to $x_6$.

\Rfig{eq-other} (a) shows the result as functions of the offset $\Dth$
for $\mu^\ast=0.97$.  In this case, a Hopf bifurcation point $P_1$ does
not appear in the plotted range $\Dth>0$.  Outside this range,
$x_2^\ast$ and $x_3^\ast$ vanish at $\Dth=0$ and violate the physical
assumptions $x_2^\ast<0$ and $x_3^\ast>0$ for $\Dth<0$.

\begin{figure}[t]
 \centering
 \includegraphics[width=\hsize, trim=0 0 5 5]{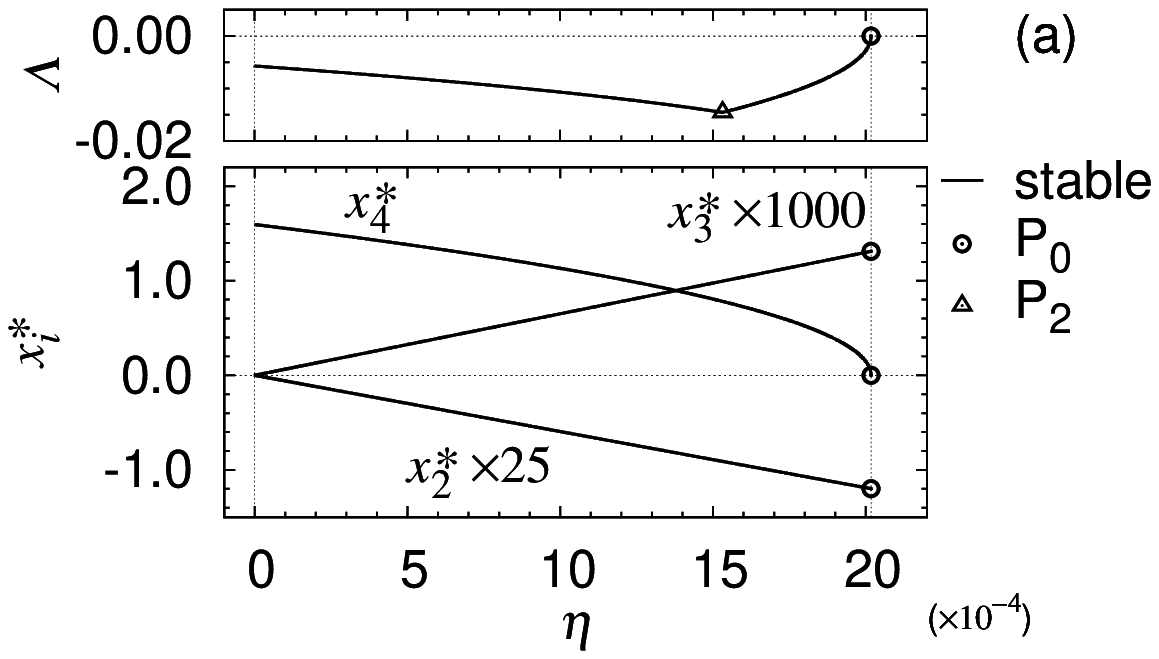}\\
 \includegraphics[width=\hsize, trim=0 0 5 5]{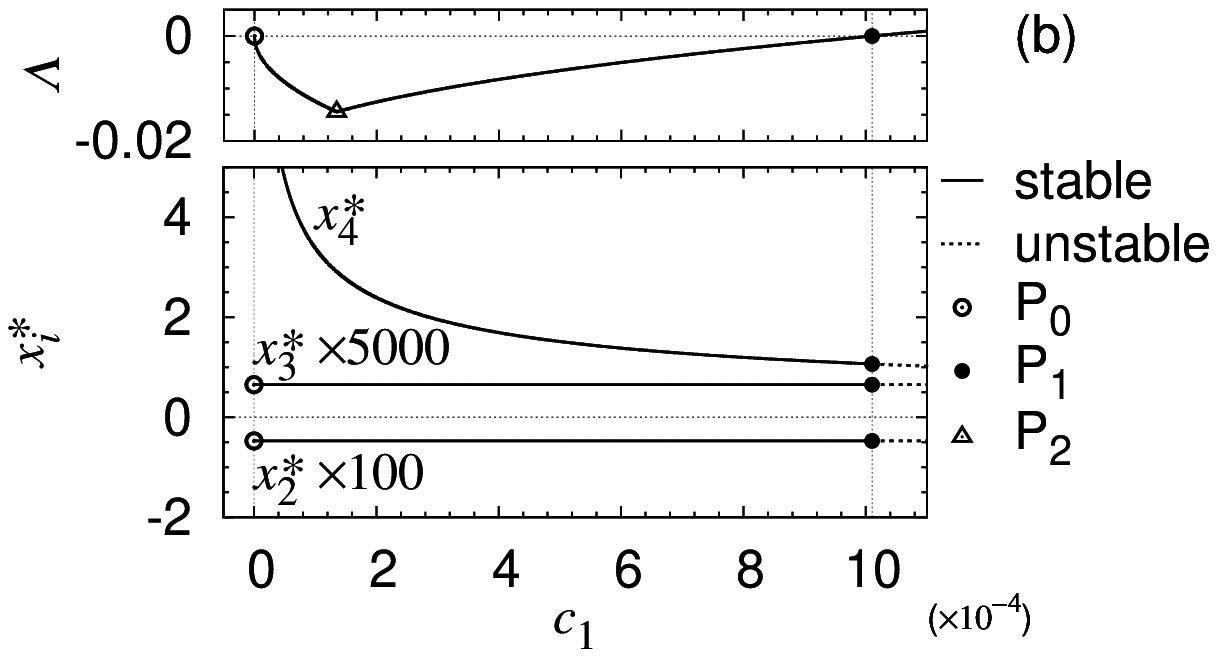}\\
 \includegraphics[width=\hsize, trim=0 0 5 5]{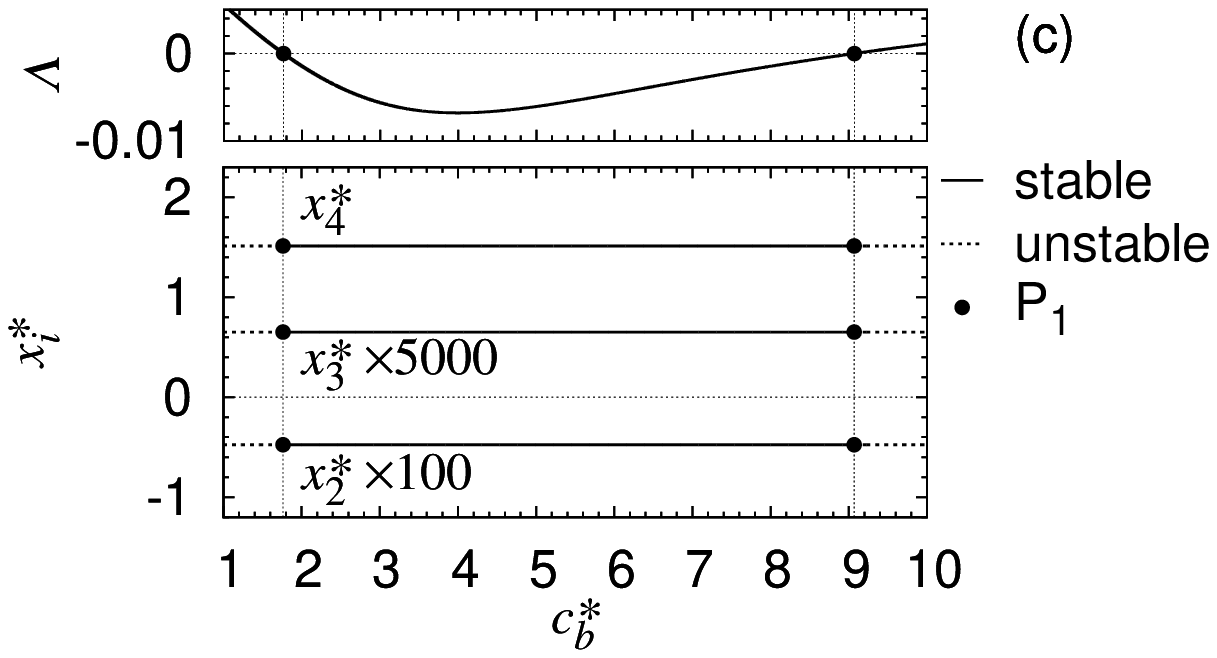}
 \caption{Maximal real part of eigenvalue $\Lam$ and steady states
 $x_i^\ast$ for $\mu^\ast=0.97$ as functions of (a) the offset $\Dth$, (b) the
 quadratic resistance $c_1$, and (c) the viscous coefficient $c_b^\ast$.}
 \lfig{eq-other}
\end{figure}

\subsection{Dependency on $c_1$ and $c_b^\ast$}
\lsec{depend2}

\Rfig{eq-other} (b) shows the results as functions of the quadratic
resistance $c_1$.
It is clear that all the critical points $P_0$, $P_1$, and $P_2$
appear, although $x_2^\ast$ and $x_3^\ast$ become constant here because
$c_1$ only affects $x_4^\ast$, as already discussed in \reqn{SSeqn}.

However, the physical results of $P_0$ are different.
That is, 
$P_0$ (or $\lammax=0$) on $\Lam(\mu^\ast)$ in \rfig{eq-mu} corresponds
to the descent velocity at rest $x_4^\ast=0$.
In contrast, $P_0$ on $\Lam(c_1)$ in \rfig{eq-other} (b) corresponds
to the infinite descent velocity $x_4^\ast\to \infty$ ($c_1\to0$).
This can be explained by the second equation in \reqn{SSeqn}, which is
hyperbolic with respect to $c_1$ and $x_4^\ast$:
\begin{equation}
 {c_1} (x_4^\ast)^2 = \relax{g^\ast Q_1 \sin (\alpha )- \mu^\ast  x_3^\ast}
 \quad=:\bar{C}>0.
 \leqn{SSeqn4}
\end{equation}
This equation exhibits the following features:
\begin{itemize}
 \item The right side of \reqn{SSeqn4} (e.g., $\bar{C}$) is expected to be
       constant because $x_3^\ast$ is determined independently of
       \reqn{SSeqn4}.
 \item The left side ${c_1}(x_4^\ast)^2$ vanishes at $P_0$ (or
       $\lammax=0$), as will be discussed in \rsec{ZQR}.
\end{itemize}
The second feature ${c_1}(x_4^\ast)^2=0$ holds when $c_1=0$ and/or
$x_4^\ast=0$.  The latter condition $x_4^\ast=0$ directly explains $P_0$
in \rfig{eq-mu} for a constant $c_1>0$. On the other hand, $P_0$ in
\rfig{eq-other} (b) can be explained by the limit $c_1\to 0$ that causes
$x_4^\ast = \sqrt{\bar{C}/c_1}\to \infty$ for the constant $\bar{C}$.
In addition, these different $P_0$ can also be explained
physically. That is, the condition ${c_1} (x_4^\ast)^2=0$ results in
vanishing of the quadratic resistance force
$c_1|x_4^\ast|x_4^\ast=0$. This can be caused by the
mechanism at rest $x_4^\ast=0$ as well as by the absence of the effect of the
quadratic resistance $c_1=0$.

\Rfig{eq-other} (c) plots the result for the nondimensional viscous
coefficient $c_b^\ast$ of the FCM.  It is found that only the Hopf
bifurcation point $P_1$ appears. Moreover, it appears that $x_i^\ast$
($i=1,2,3$) are all constant because the steady-state equation
in \reqn{SSeqn} is independent of $c_b^\ast$, which only affects the
components of the Jacobian matrix as $\mp c_b^\ast\mu^\ast$ and
$-c_b^\ast$ in \reqn{Dtt}.

\section{Stability limits}

Finally, we numerically continue the critical points in two-parameter planes
to characterize the stability limits of the FCWIP model.

\subsection{Conditions of the critical points}

\subsubsection{Zero quadratic resistance}
\lsec{ZQR}

The condition of $P_0$ (or $\lammax=0$) is mathematically equivalent to
$a_5=0$ in the eigenvalue equation \reqn{h}. It follows that
\begin{align}
 &0=a_5 
  =\det\mm^{-1} \cdot 2 c_1 x_4^\ast g^\ast 
 \notag\\&\qquad \qquad \times
  \Big\{-C_1^\ast Q_4 -C_2^\ast Q_3 \{C_1^\ast g^\ast Q_4 -(1+\mu^\ast)\}\Big\}
 \notag\\
 & \Longleftrightarrow\;
 0= c_1 x_4^\ast 
 \Big\{C_1^\ast Q_4 + C_2^\ast Q_3 \{C_1^\ast g^\ast Q_4 - (1+\mu^\ast)\}\Big\}
 \notag\\
 & \Longleftarrow\;\, c_1 x_4^\ast=0 
  \notag\\ &
 \Longleftrightarrow\;
 c_1=0\;\text{or}\;x_4^\ast=0 
 \;\Longleftrightarrow\;
 c_1 (x_4^\ast)^2 =0.
\end{align}
Therefore, it is mathematically shown that the sufficient condition for
$P_0$ (or $\lammax=0$) is given by the zero quadratic
resistance $c_1(x_4^\ast)^2 = 0$.

As discussed in \rsec{depend2}, the
condition $P_0$ (or $c_1(x_4^\ast)^2 = 0$) causes two distinct
descent velocities: $x_4^\ast=0$ for $c_1>0$ and $x_4^\ast=\infty$ for
$c_1=0$. Therefore, we denote them as $P_0^0$ and $P_0^\infty$,
respectively, in the following sections.

The condition of $P_0^0$ is given by \reqn{SSeqn4} with $c_1(x_4^\ast)^2
= 0$, from which we can eliminate $x_2^\ast$ and $x_3^\ast$ to obtain
\begin{multline}
 \Phi_0 := \eta + \arcsin\left(\frac{Q_1\sin\alpha}{Q_3\mu^\ast}\right)
  +\frac{g^\ast Q_1\sin\alpha}{\mu^\ast}
 \\
  -\arcsin\left(\frac{Q_1(1+\mu^\ast)\sin\alpha}{Q_4\mu^\ast}\right)
  =0.
 \leqn{P0}
\end{multline}
On the other hand, $P_0^\infty$ is simply given by $c_1=0$.  Because the
condition \reqn{P0} does not contain $x_2^\ast$ and $x_3^\ast$, the zero
quadratic resistance points $P_0^0$ and $P_0^\infty$ are determined
independently of the pendulum angles $x_2^\ast$ and $x_3^\ast$.

\subsubsection{Hopf bifurcation}
\lsec{Hopf}

The condition $P_1$ (or $\lammax=\pm j\Omega$) for a Hopf
bifurcation point is given by $\Re h(j\Omega) = \Im h(j\Omega) = 0$.
On eliminating $\Omega$ from them, we obtain
\begin{multline}
\Phi_1 := a_5 \Big\{a_4 (-a_1 a_2 a_3 + a_3^2 + a_1^2 a_4) 
\\
+ \big(-a_2 a_3 + a_1 (a_2^2 - 2 a_4)\big) a_5 + a_5^2\Big\} = 0
\leqn{P1}
\end{multline}
for the Hopf bifurcation point. It is implied from \reqn{P1} that $P_1$
coincides with $P_0^0$ (or $P_0^\infty$) because $a_5=0$ for $P_0^0$ (or
$P_0^\infty$) leads to $\Phi_1=0$.

Note that, rigorously speaking, the above condition provides only a
necessary condition of a Hopf bifurcation point; however, it leads to
satisfactory results in the present analysis.

\subsubsection{Minimal time constant}

We numerically detect the condition of $P_2$ for the minimal time
constant that satisfies
\begin{equation}
 |s_0 - s_1| < 10^{-9}
 \leqn{P2}
\end{equation}
where $s_0$ is the maximal real eigenvalue and $s_1$ is the real part of
the complex conjugate eigenvalues, as defined in \reqn{5eig}.
In numerical calculations, the parameter values considered are swept to
detect a point that satisfies \reqn{P2}, where the point in the first
detection is taken as the point detected.

Note that we attempted to derive an equation for the minimal time constant
in a closed form of $a_1,\cdots,a_5$ based on a given form of eigenvalue
equation:
\begin{math}
 h(\lam) = (\lam - \lammax) (\lam - \lammax - jv) 
 (\lam - \lammax + jv)(\lam^2 + p \lam + q)
\end{math},
however, the result was very weak to detect $P_2$ precisely. Therefore,
in this paper, we employ the numerical method mentioned above, although
another approach would be possible for an analytical expression of
$P_2$.

\subsection{Numerical continuation of the critical points}

\Rfig{bif} plots the sets of the critical points $P_0^0$, $P_1$, and
$P_2$ on two-parameter planes obtained from the numerical solutions of
\reqn{P0}, \reqn{P1}, and \reqn{P2} under \reqn{SSeqn}.
The results on the $(\mu^\ast,\Dth)$, $(\mu^\ast,c_1)$, and
$(\mu^\ast,c_b^\ast)$ planes are labeled (a), (b), and (c) respectively in
\rfig{bif}.
The  solid line denotes the set of the zero quadratic
resistance point $P_0^0$ for $x_4^\ast=0$, the  dotted line
denotes the set of the Hopf bifurcation point $P_1$, and the 
chained line denotes the set of the minimal time constant point
$P_2$. The plots of $P_0^\infty$ for $c_1=0$ do not appear in the ranges
of these plots. The hatched areas represent the stable regions of the
steady state satisfying $\Lam < 0$ in \reqn{Lam}.

\begin{figure}[t]
 \centering
 \includegraphics[width=\hsize, trim=0 0 0 0]{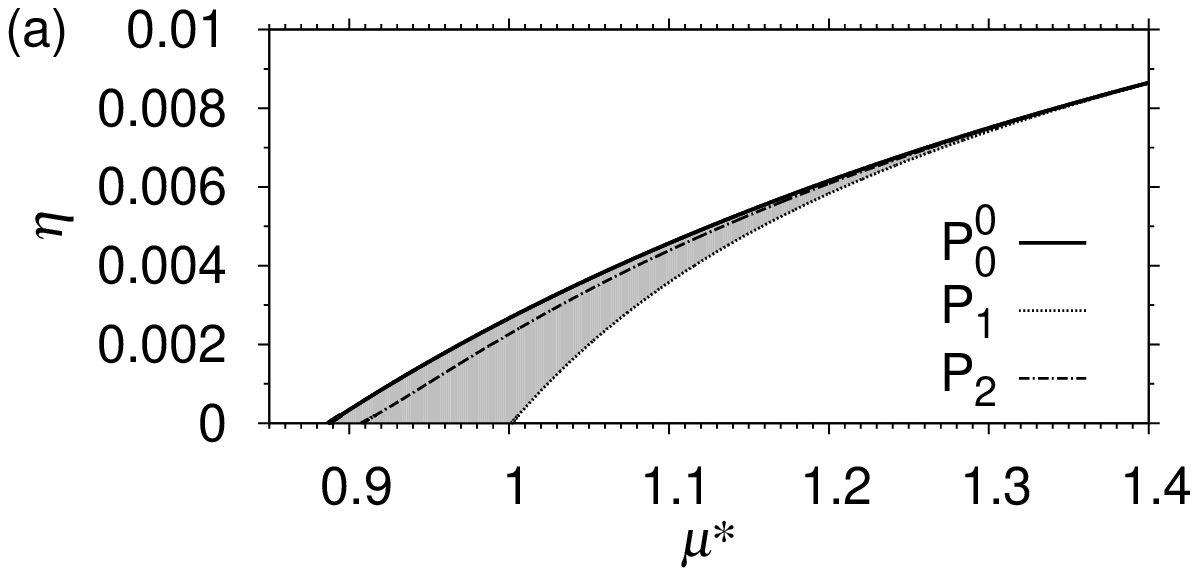}\\
 \includegraphics[width=\hsize, trim=0 0 0 0]{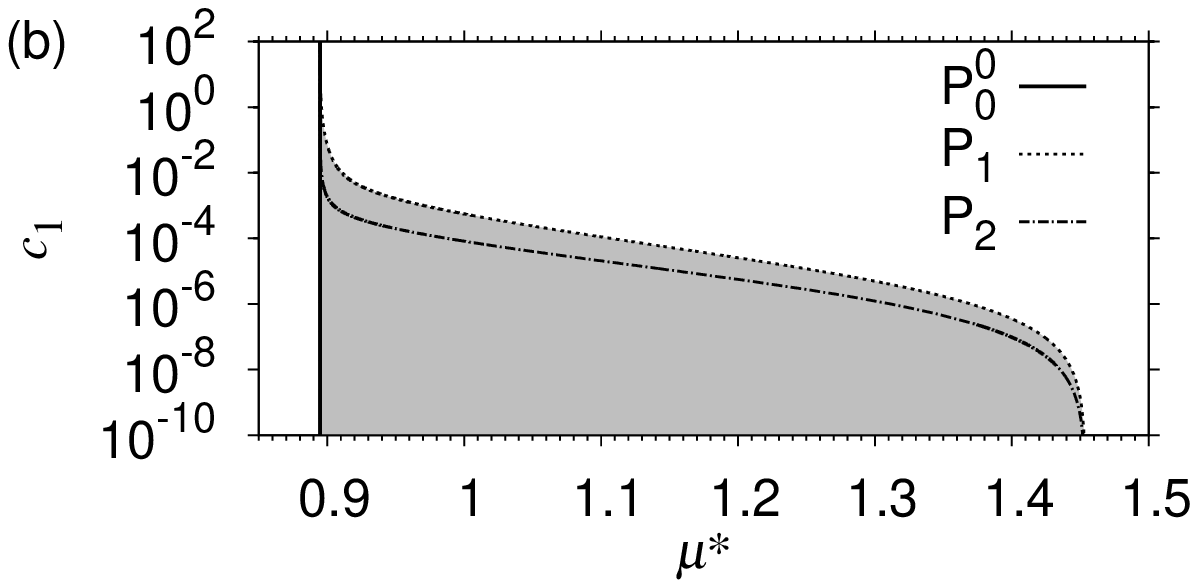}\\
 \includegraphics[width=\hsize, trim=0 0 0 0]{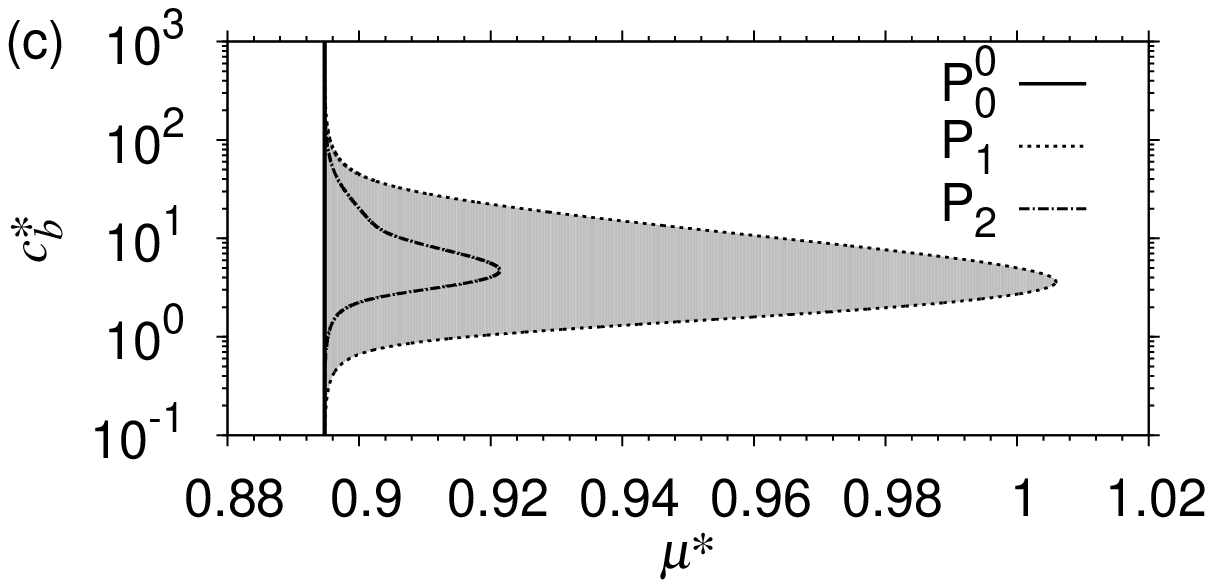}
 \caption{Continuation of the critical points $P_0^0$, $P_1$, and $P_2$
 on the two-parameter planes: (a) for the $(\mu^\ast,\Dth)$ plane, (b) for
 the $(\mu^\ast,c_1)$ plane, and (c) for the
 $(\mu^\ast,c_b^\ast)$ plane. The hatched areas represent
 asymptotically stable conditions of the steady states. }
 \lfig{bif}
\end{figure}

It is clearly seen in \rfig{bif} that the stable regions are
bounded by $P_0^0$ and $P_1$ and that the minimal time constant point
$P_2$ is sandwiched between them. Note that in \rfig{bif} (a), the
plots are bounded by the assumption $\Dth>0$ and that in \rfig{bif} (b)
and (c), $P_0^0$ lies along the vertical line at
$\mu^\ast=\bar{\mu}^\ast\approx 0.89474$. This is because $P_0^0$ is determined
independently of $c_1$ and $c_b^\ast$ in \reqn{P0}.
Moreover, it appears that the critical points $P_0^0$, $P_1$, and $P_2$ tend
to coincide as $\mu^\ast$ increases on the $(\mu^\ast,\Dth)$ plane, as
$c_b^\ast$ increases and decreases on the $(\mu^\ast,c_b^\ast)$ plane,
and as $c_1$ increases on the $(\mu^\ast,c_1)$ plane.  In contrast, as
$c_1$ decreases on the $(\mu^\ast,c_1)$ plane, $P_1$ and $P_2$ also tend to
coincide but they approach $c_1=0$ or $P_0^\infty$ instead of $P_0^0$.
As discussed in \rsec{Hopf}, the convergence between $P_0^0$ (or
$P_0^\infty$) and $P_1$ can be explained by \reqn{P1}, 
in which the condition $\Phi_1=0$ for the Hopf bifurcation point $P_1$
contains $a_5=0$ for the zero quadratic resistance point $P_0^0$ (or
$P_0^\infty$).
As shown in \rfig{bif-im}, however, it can be numerically proven that
a purely imaginary eigenvalue $\lammax=j\Omega$ along $P_1$ does not
vanish even when $P_1$ coincides with $P_0^0$ (or $P_0^\infty$) in the
parameter planes.  Therefore, the Hopf bifurcation point $P_1$ does not
degenerate and double-zero eigenvalues never arise at that point.

From an engineering viewpoint, the results obtained above imply
that the FCWIP model allows a certain amount of tolerance in
parameter settings because the stable conditions are obtained as
simply connected finite areas.  This suggests the possibility that the
proposed mechanism can work even if there are some manufacturing
errors.  Furthermore, the stable area on the
$(\mu^\ast,\Dth)$ plane in \rfig{bif} (a) forms a monotonically
increasing shape, which implies that the offset $\Dth$ can be designed
to shift the stable range of the friction coefficient $\mu^\ast$.

\begin{figure}[t]
 \centering
 \includegraphics[width=\hsize, trim=0 0 0 0]{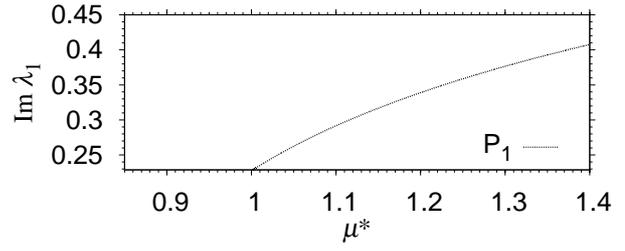}
 \caption{Imaginary part of complex conjugate eigenvalues along $P_1$
 in \rfig{bif} (a).}
 \lfig{bif-im}
\end{figure}

\subsection{Numerical evaluation of the descent velocity}

In view of engineering applications, the descent velocity (or angular
velocity of the wheel) $x_4^\ast$ must be adjusted to a value suitable
for the intended use. \Rfig{bif-om} shows the values of $x_4^\ast$
mapped into a gray scale within the stable areas on the parameter
planes in \rfig{bif}. $x_4^\ast$ values are numerically obtained by
solving \reqn{SSeqn}.

It is clear from \rfig{bif-om} that $x_4^\ast$ tends toward zero as the
parameter conditions approach $P_0^0$ (solid lines), which
is in agreement with the definition of $P_0^0$. Especially, in
\rfig{bif-om} (b), it is also clarified that $x_4^\ast$ is diverging as
the condition approaches $P_0^\infty$ or $c_1=0$. Moreover, it appears
that $x_4^\ast$ changes smoothly and monotonically with the variations
of the parameters. This suggests that one can adjust the descent
velocity $x_4^\ast$ by continuously shifting the parameters. However, it
is also observed in \rfig{bif-om} (b) that a sufficiently large value of
the quadratic resistance $c_1$ is required to stabilize the mechanism
because although a range of $\mu^\ast$ exists for small $x_4^\ast$ near
$\mu^\ast=\bar{\mu}^\ast\approx 0.89474$, it narrows significantly as
$c_1$ decreases.
In addition, it appears in \rfig{bif-om} (c) that $x_4^\ast$
does not depend on $c_b^\ast$, as discussed in the last paragraph
in \rsec{linear}.

\begin{figure}[t]
 \centering
 \includegraphics[width=\hsize, trim=0 10 0 10]{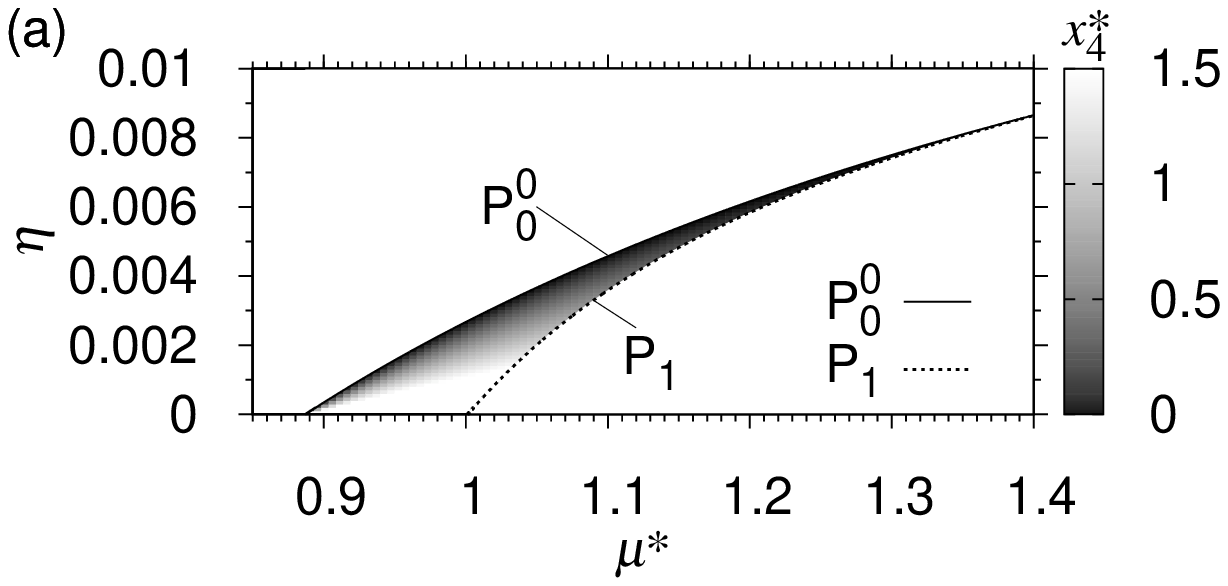}\\
 \includegraphics[width=\hsize, trim=0 10 0 10]{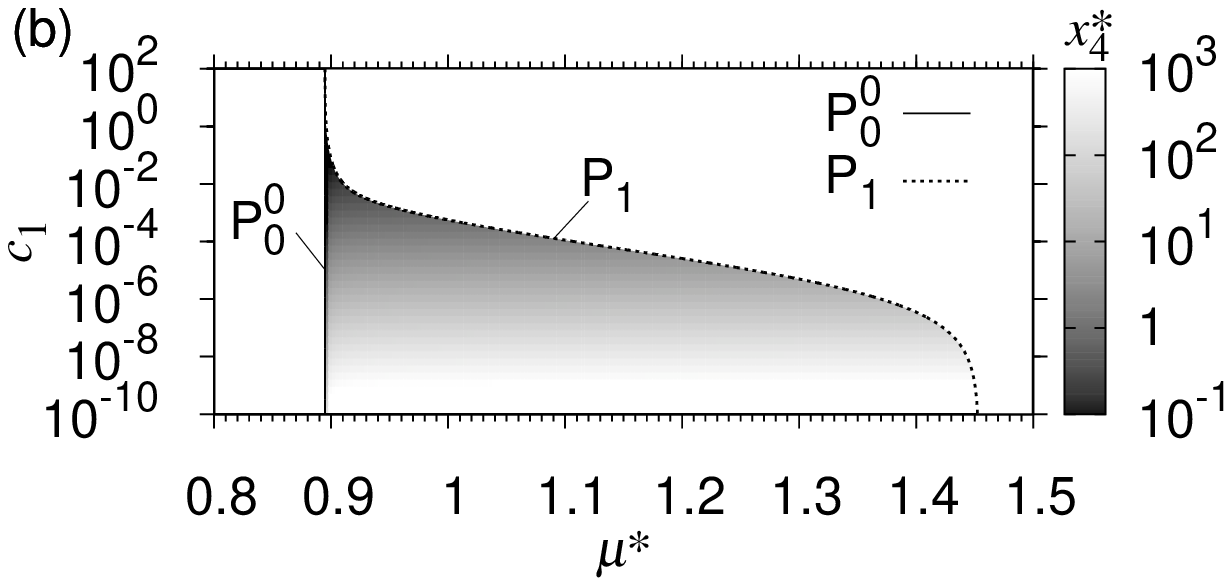}\\
 \includegraphics[width=\hsize, trim=0 10 0 10]{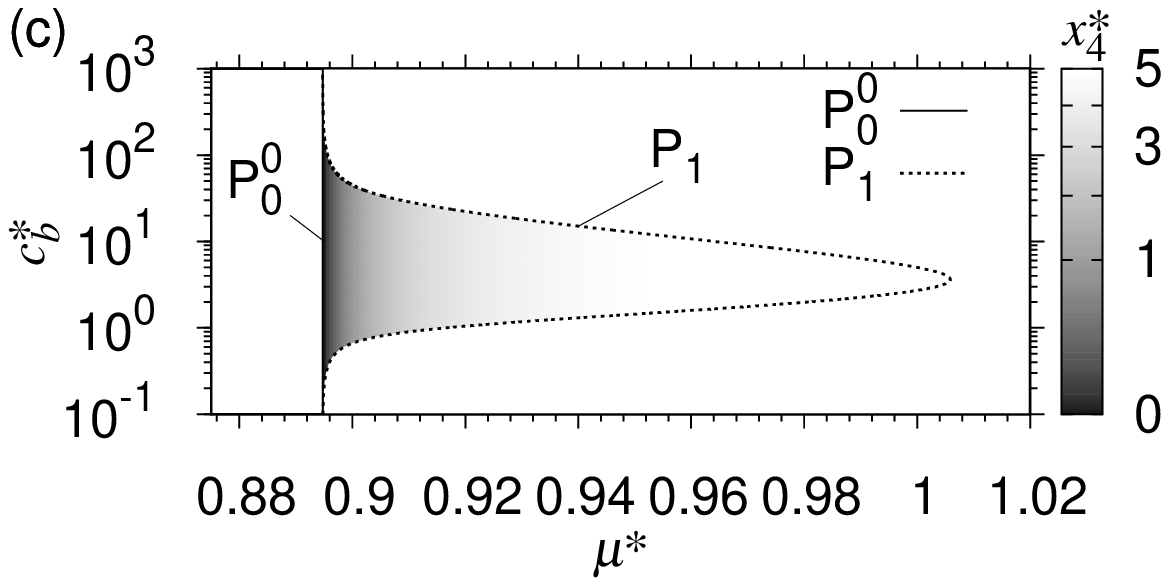}
 \caption{Angular velocities of the wheel $x_4^\ast$ within the stable areas 
 on the two-parameter planes: (a) for the $(\mu^\ast,\Dth)$ plane; (b) for
 the $(\mu^\ast,c_1)$ plane; and (c) for the
 $(\mu^\ast,c_b^\ast)$ plane. }
 \lfig{bif-om}
\end{figure}

\section{Conclusion}

For a non-electrified method to stabilize a wheeled inverted
pendulum descending a slope, we proposed the FCWIP mechanism, a
wheeled double pendulum,  whose second pendulum transforms gravity
force into brake force acting on the wheel. We conducted
steady-state analysis of the proposed model and obtained the following results:
\begin{itemize}
 \item The steady angles $(x_2^\ast,x_3^\ast)$ depend on $(\mu^\ast, \eta)$.
 \item The steady descent velocity $x_4^\ast$ depends on $(\mu^\ast,
       \eta, c_1)$.
 \item The stability depends on $(\mu^\ast, \eta, c_1, c_b^\ast)$.
\end{itemize}
Then, we found three types of critical points in the steady states, as
\begin{itemize}
 \item $P_0$: the point for zero quadratic resistance on the wheel.
       ($P_0^0$ for $x_4^\ast=0$ and $P_0^\infty$ for $x_4^\ast=\infty$
       in detail)
 \item $P_1$: the Hopf bifurcation point.
 \item $P_2$: the point for the minimal time constant.
\end{itemize}

Finally, we conducted numerical continuations of these points on 
the two-parameter planes and evaluated the descent velocity to obtain the
following results:
\begin{itemize}
 \item The stable conditions are obtained as simply connected finite areas
       on the parameter planes, bounded by $P_0^0$ and $P_1$.
 \item The minimal time constant point $P_2$ is sandwiched between $P_0^0$ and
       $P_1$.
 \item The descent velocity $x_4^\ast$ changes smoothly and
       monotonically with the parameter variations.
\end{itemize}

The abovementioned results lead to the conclusion that the
parameter selection to design the FCWIP mechanism stabilized on a slope
will not be highly sensitive, at least in theory.

In future work, we plan to develop a physical FCWIP mechanism. For
this purpose, we will introduce stick-slip effects into the friction
term in our model and investigate the effect on the
stabilities. We also plan to conduct stochastic analysis on
the FCWIP model to consider robustness against random disturbances and
random parameter fluctuations.

\begin{acknowledgements}
This work was supported in part by grants from the ZERO Design Project of
Utsunomiya University (a proj\-ect seeking novel engineering designs
robust against massive outages of power, water, logistics, etc.).

\end{acknowledgements}

%\bibliographystyle{NonlinearDynamics}
%\bibliography{FCWIP,PassiveStabilization}

\begin{thebibliography}{10}
\providecommand{\url}[1]{{#1}}
\providecommand{\urlprefix}{URL }
\expandafter\ifx\csname urlstyle\endcsname\relax
  \providecommand{\doi}[1]{DOI~\discretionary{}{}{}#1}\else
  \providecommand{\doi}{DOI~\discretionary{}{}{}\begingroup
  \urlstyle{rm}\Url}\fi

\bibitem{McGeer1990}
McGeer, T.: {Passive Dynamic Walking}.
\newblock The International Journal of Robotics Research \textbf{9}(2), 62--82
  (1990)

\bibitem{Ikemata2009}
Ikemata, Y., Sano, A., Yasuhara, K., Fujimoto, H.: {Dynamic effects of arc feet
  on the leg motion of passive walker}.
\newblock In: 2009 IEEE International Conference on Robotics and Automation,
  pp. 2755--2760. IEEE (2009)

\bibitem{Coleman1998}
Coleman, M., Ruina, A.: {An Uncontrolled Walking Toy That Cannot Stand Still}.
\newblock Physical Review Letters \textbf{80}(16), 3658--3661 (1998)

\bibitem{Garcia1998}
Garcia, M., Chatterjee, A., Ruina, A., Coleman, M.: {The simplest walking
  model: stability, complexity, and scaling.}
\newblock Journal of biomechanical engineering \textbf{120}(2), 281--288 (1998)

\bibitem{Goswami1998}
Goswami, A., Thuilot, B., Espiau, B.: {A study of the passive gait of a
  compass-like biped robot symmetry and chaos}.
\newblock The International Journal of Robotics Research \textbf{17}(12),
  1282--1301 (1998)

\bibitem{HIRATA2011}
Hirata, K.: {On Internal stabilizing mechanism of passive dynamic walking}.
\newblock SICE Journal of Control, Measurement, and System Integration
  \textbf{4}(1), 29--36 (2011)

%==>passive
\bibitem{BLACK1964}
Black, H. D.: {A passive system for determining the attitude of a satellite}.
\newblock AIAA Journal \textbf{2}(7), 1350--1351 (1964)

\bibitem{FISCHELL1964}
Fischell, R. E.: {Passive Gravity-Gradient Stabilization for Earth Satellites}.
\newblock {Applied Mathematics and Mechanics}, vol.~7.
\newblock Elsevier, 13--30 (1964)

\bibitem{He1999}
He, C., Liu, G., Yang, L., Tian, Y.: {On the passive stabilization of the
  equilibrium state of Lagrangian systems}.
\newblock Acta Mechanica \textbf{134}(1-2), 17--26 (1999)

\bibitem{Peiffer2000}
Peiffer, K., Savchenko, A.: {On Passive Stabilization in Critical Cases}.
\newblock Journal of Mathematical Analysis and Applications \textbf{244}(1),
  106--119 (2000)

%<==passive

\bibitem{Ulrich2005} %7
Ulrich, K.T.: {Estimating the technology frontier for personal electric
  vehicles}.
\newblock Transportation Research Part C: Emerging Technologies
  \textbf{13}(5-6), 448--462 (2005)

\bibitem{Pathak2005}
Pathak, K., Franch, J., Agrawal, S.K.: {Velocity and Position Control of a
  Wheeled Inverted Pendulum by Partial Feedback Linearization}.
\newblock IEEE Transactions on Robotics \textbf{21}(3), 505--513 (2005)

\bibitem{Kim2006}
Kim, Y., Kim, S.H., Kwak, Y.K.: {Dynamic Analysis of a Nonholonomic Two-Wheeled
  Inverted Pendulum Robot}.
\newblock Journal of Intelligent and Robotic Systems \textbf{44}(1), 25--46
  (2006)

\bibitem{Huang2010} %8
Huang, J., Zhi-Hong, G., Matsuno, T., Fukuda, T., Sekiyama, K.: {Sliding-Mode
  Velocity Control of Mobile-Wheeled Inverted-Pendulum Systems}.
\newblock IEEE Transactions on Robotics \textbf{26}(4), 750--758 (2010)

\bibitem{Su2010}
Su, K., Chen, Y., Su, S.: {Design of neural-fuzzy-based controller for two
  autonomously driven wheeled robot}.
\newblock Neurocomputing \textbf{73}(13-15), 2478--2488 (2010)

\bibitem{Vermeiren2011}
Vermeiren, L., Dequidt, A., Guerra, T.M., Rago-Tirmant, H., Parent, M.:
  {Modeling, control and experimental verification on a two-wheeled vehicle
  with free inclination: An urban transportation system}.
\newblock Control Engineering Practice \textbf{19}(7), 744--756 (2011)

\end{thebibliography}

\end{document}